\def\etal{{\it et al.\thinspace}}
\def\eg{{\it e.g.\ }}

\def\ie{{\it i.e.\ }}

\def\gsim{~\rlap{$>$}{\lower 1.0ex\hbox{$\sim$}}}
\def\lsim{~\rlap{$<$}{\lower 1.0ex\hbox{$\sim$}}}

\def\idm#1{{\mbox{\scriptsize #1}}}

\newcommand\Chi{{(\chi^2_\nu)^{1/2}}}
\def\astrobj#1{#1\ }

\newcommand\pstar{{\astrobj{HD~74156}}}

\documentclass[12pt,preprint]{aastex}
\begin{document}

\title{The Successful Prediction of the Extrasolar Planet HD 74156 d}

\author{Rory Barnes\altaffilmark{1}, Krzysztof Go\'zdziewski\altaffilmark{2}, Sean N. Raymond\altaffilmark{3,4}}

\altaffiltext{1}{Lunar and Planetary Laboratory, University of Arizona,
Tucson, AZ 85721, rory@lpl.arizona.edu}
\altaffiltext{2}{Toru\'n Centre for Astronomy, N. Copernicus University, Toru\'n, Poland}
\altaffiltext{3}{NASA Postdoctoral Program Fellow}
\altaffiltext{4}{Center for Astrophysics and Space Astronomy, University of Colorado, Boulder, CO 80309}


\begin{abstract}
Most of the first-discovered extrasolar multi-planet systems were found to lie
close to dynamically unstable configurations. However a few observed
multi-planet systems (\eg HD 74156) did not show this trait. Those systems could share this property if they contain an additional planet in between those that are
known.  Previous investigations identified the properties of hypothetical planets that would place
these systems near instability. The hypothetical planet in HD 74156 was
expected to have a mass about equal to that of Saturn, a semi-major axis
between 0.9 and 1.4 AU, and an eccentricity less than 0.2. HD 74156 d, a
planet with a mass of 1.3 Saturn masses at 1.04 AU with an eccentricity of
0.25, was recently announced.
We have reanalyzed all published data on this system in order to
place tighter constraints on the properties of the new planet. We find two
possible orbits for this planet, one close to that already identified and
another (with a slightly better fit to the data) at $\sim 0.89$ AU. We
also review the current status of other planet predictions, discuss the
observed single planet systems, and suggest other systems which may
contain planets in between those that are already known. The confirmation
of the existence of HD 74156 d suggests that planet formation is an
efficient process, and planetary systems should typically contain many
planets.
\end{abstract}

\section{Introduction}

The detection of extrasolar planets has provided an unprecedented
opportunity to test models of planet formation. One such model is the
``Packed Planetary Systems'' (PPS) hypothesis (Barnes \& Raymond 2004 (hereafter RB04); Raymond \& Barnes 2005 (hereafter RB05); Raymond \etal
2006; Barnes \& Greenberg 2007; see also Barnes \& Quinn 2001, 2004)
which posits that, between the innermost and outermost giant planets,
planetary systems formed such that they were filled to capacity; an
additional planet would create an unstable system. Other
investigations gave consistent results (\eg Rivera \& Lissauer 2000;
Bois \etal 2003; Go\'zdziewski \etal 2006). These analyses were
grounded in considerable theoretical work to examine the (in)stability
of planetary systems (\eg Chambers \etal 1996; Weidenschilling \&
Marzari 1996; Ida \& Lin 1997). 

The PPS hypothesis was suggested because five of the first six known
extrasolar multi-planet systems showed evidence for packing:
$\upsilon$ And (Butler \etal 1999), 47 UMa (Fischer \etal 2002), GJ
876 (Marcy \etal 2001b), HD 82943 (Mayor \etal 2004), and HD 168443
(Marcy \etal 2001a) were packed (Barnes \& Quinn 2001, 2004; BR04),
but HD 74156 (Naef \etal 2004) was not. This high frequency of packing
suggests that it must be a common feature of planetary systems. The
PPS hypothesis takes this suggestion a step further and proposes that
all planetary systems have tended to form dynamically
full.

For the PPS model to be correct, then seemingly non-packed systems
require at least one additional planet to fill them up. This
conjecture led BR04 and RB05 to search for stable regions in between
known planets in several systems, including HD 74156. These
investigations presumed that, by identifying a stable region in
between the known planets, it would be possible to predict the mass
and orbit of a previously unknown planet. BR04 explained the reasons to
expect additional planets and outlined how to predict planets
through numerical simulations of test particles. RB05 then placed
massive companions in the gaps identified in BR04 in order to predict
the likely orbital elements of planets that could be detected. Note
that BR04 used the initially reported orbital elements and found HD
74156 was packed. RB05 used the revised orbits, published in Naef
\etal (2004), and found the system was not packed.

An additional planet would require a mass and orbit that guaranteed
dynamical stability for the age of the system. The greatest difficulty
with such an analysis is that the size of the stable zone and the mass
of the hypothetical planet are degenerate; lower mass planets have a
wider stable zone, larger planets have a narrower zone. In order to
estimate the mass, RB05 considered the mass of the two known
planets. As both these planets are larger than Jupiter (2 and 8
Jupiter masses), then a planet in between the two would also need to
be large in order for the protoplanetary disk to have a plausible
surface density profile (\eg a power law). At the same time, the mass
would have to be small enough to have avoided detection by the initial
observations. In order to help quantify the predicted mass of the
hypothetical planets, RB05 considered Saturn-mass, Jupiter-mass and 10
Jupiter-mass objects. They found the HD 74156 system had an 80\%
probability of being able to support a Saturn-mass planet, a 40\%
probability for a Jupiter-mass planet, and a 10\% probability of 10
Jupiter-mass planet. The larger masses would likely have been detected
previously, so RB05 settled on a Saturn mass as the likely mass of the
putative companion.

Bean \etal (2008) discovered HD 74156 d, an extra-solar planet with a mass
40\% greater than Saturn, at 1.04 AU with an eccentricity of
0.25. This planet is consistent with the PPS model: RB05 predicted a
Saturn-mass planet with semi-major axis, $a$, in the range $0.9 \lsim
a \lsim 1.4$ AU and eccentricity, $e$, in the range $0 \le e \lsim
0.2$. Here we reanalyze all available data for HD 74156
($\S$ 2), discuss the current status of the PPS model in
$\S$ 3, and draw general conclusions in $\S$ 4.

\section{The Orbit of HD 74156 d}
The best-fit orbits published by Bean \etal (2008) for the planets
orbiting HD 74156 are actually unstable on $\sim 10^5$ year
timescales. The instability arises due to strong gravitational
interactions resulting from close approaches between the inner two
planets, b and d. In this section we consider all published
observations and identify four plausible stable fits to the data.

Currently, all published radial velocity (RV) data for \pstar{} appear
in papers by Naef \etal (2004) (51 observations gathered with the
ELODIE spectrometer with a mean dispersion of 12.7~ms$^{-1}$, and 44
measurements of CORALIE with a mean dispersion of 8.5~ms$^{-1}$), and
in Bean \etal (2008), who published 82 precision measurements from the
Hobby-Eberly Telescope (HET), with mean dispersion of
2.7~ms$^{-1}$. The combined data set covers $\sim 9.33$~yr. We binned
a few measurements in the CORALIE and ELODIE data sets which were done
during one night. We also shifted ELODIE and CORALIE observations with
respect to the mean RV in both set.  We rescaled data
errors by adding jitter of $4$~ms$^{-1}$ in quadrature. In all
calculations, we adopted the mass of the parent star to be
$1.24$~M$_{\sun}$, following Naef \etal (2004). This procedure is
slightly different than (but consistent with) that in Bean \etal (2008).

First, we searched for the best Keplerian fits to the combined data
set with the hybrid optimization code (Go\'zdziewski \& Migaszewski
2006) relying on the genetic algorithm (Charbonneau 1995).
Because the number of planets in the system is unknown, we considered
models with two, three and four planets. The so-called velocity
offsets of the telescopes are the free parameters in the model. Hence,
the $N$-planet model depends on $5N+3$ primary parameters, \ie tuples
($K,P,e,\omega,\tau$) comprising of velocity amplitude $K$ in
ms$^{-1}$, orbital period $P$ in days, eccentricity, longitude of
pericenter $\omega$ in degrees, and time of periastron passage $\tau$
in JD-245000, for each planet, respectively, as well as the
offsets. The fit quality is measured in terms of reduced $\Chi$ and an
rms.

The best fit to the two-planet model yields $\Chi = 1.63$ and an rms
$= 11.5$~ms$^{-1}$. The four-planet model could not converge on any
solution in which the RV contribution from all planets exceeded the
mean uncertainty of the measurements (assuming detectable masses), and
therefore could not produce any reasonable fit to the data. However,
as we will see below, the three-planet model gives statistically
better fits to the data, so we will focus on it.

We found the best three-planet fit yields $\Chi = 1.28$ and drops
the rms to $= 9.4$~ms$^{-1}$, which represent a significant
improvement over those of the two-planet fit. The three-planet fit,
given in terms of parameter tuples is the following: (113.296, 51.641,
0.640, 175.318, 1671.231) for planet~b, (13.226, 277.378, 0.351,
294.266, 4221.105), for the new planet~d, and (108.5127, 2433.177,
0.331, 275.455, 3497.965) for planet~c. The velocity offsets are
$(-2.94, 3.66, -6.01)$~ms$^{-1}$ for ELODIE, CORALIE and HET,
respectively. Curiously, the period of the new planet~d is $\sim
277$~days, and is significantly different from the best-fit solution
quoted in Bean et al. (2008), who report $P_d = 347$~days. This
discrepancy most likely results from our different handling of the
errors, as well as the inclusion of 10 additional data
points. However, we also find a similar local minimum at $P_d =
349$~days (close to the fit in Bean \etal 2008), as well as $P_d =
900$~days, but with large $e_d$ (0.6).

Because $e_{\idm{d}}$ can reach large values in the Keplerian fit we
cannot assume that the best-fit configurations are dynamically
stable. To account for the mutual interactions and test for stability,
the ensemble of Keplerian solutions has been used as initial
conditions in an $N$-body model (Laughlin \& Chambers 2001). The
Keplerian fits are not a good representation of the system because
they lead to significant degradation of $\Chi$. Hence, we refined
these Keplerian fits to account for the mutual interactions. In
addition, we tested formal stability (as understood through chaotic or
quasi-periodic character of orbital configurations) of these refined
solutions with MEGNO (Cincotta \& Sim\'o 2000; Cincotta, Giordano \&
Sim\'o 2003; Go\'zdziewski \etal 2001). MEGNO has been computed over
$3\cdot 10^4$ orbital periods of the outermost planet which is long
enough to account for the destabilizing influence of short-term mean
motion resonances.

\begin{figure}
\includegraphics{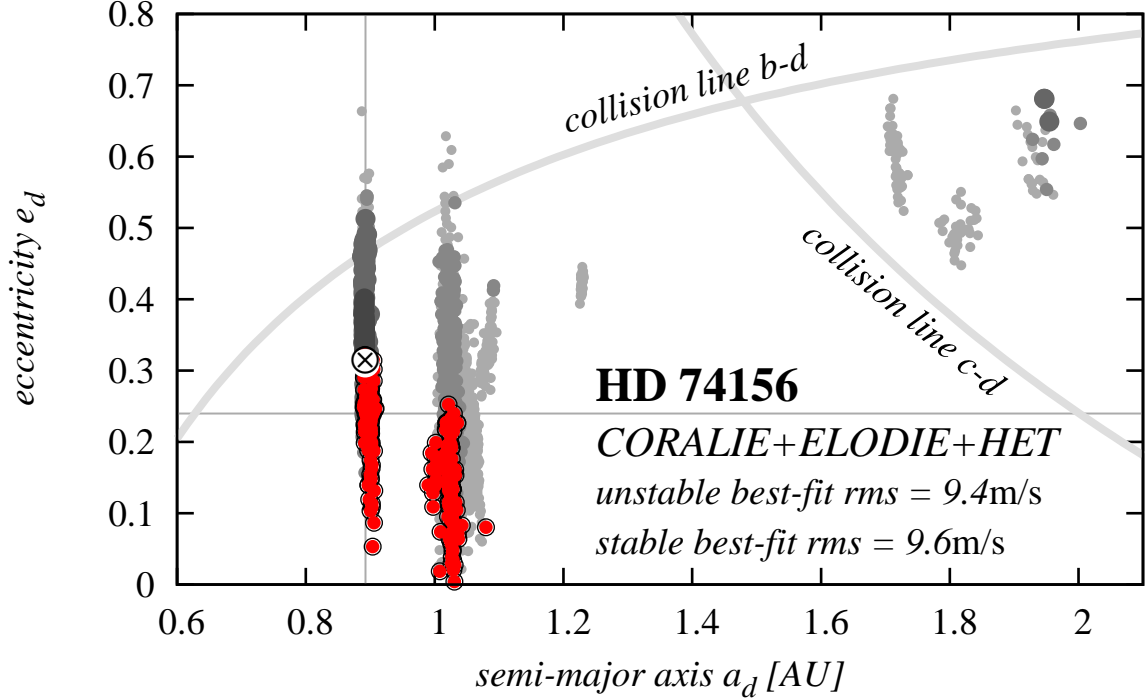}
\caption{Fit parameters of the coplanar Newtonian model to the data published in
Naef \etal (2004)  and Bean \etal (2008).  Osculating elements at the epoch of the first observation are projected onto the $(a_{\idm{d}},e_{\idm{d}})$-plane. Gray curves are the collision lines of the orbits computed with the elements of the
innermost and the outermost companions fixed at their best fit values. The best fit solution (Fit 1) is marked by the intersection of the horizontal and vertical lines. For reference, the best fit (unstable) configuration is marked with a white crossed circle. Light-gray circles are for fits with$\Chi<1.35$ (an rms limit of $\sim 10$~ms$^{-1}$). Darker  gray circles are for
solutions with  $\Chi<1.33$ and $\Chi<1.31$, respectively. Stable fits are
marked with red circles.  }
\label{fig:fit}
\end{figure}

\begin{figure}
\includegraphics{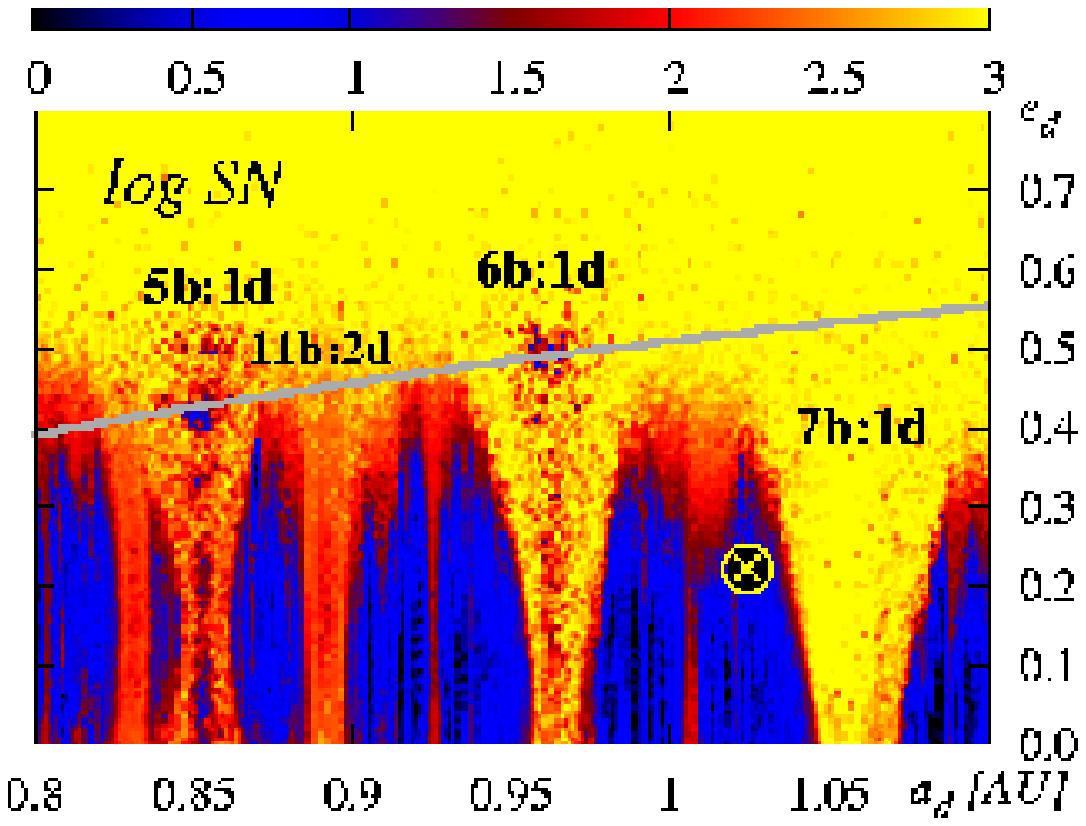}
\caption{The dynamical map for configurations near Fit 2 in terms of the Spectral Number
$\log (SN)$ (Michtchenko \& Ferrraz-Mello 2001). The character of the solutions is color-coded: yellow means chaotic systems
and black means regular systems. Fit 2 is marked with a crossed
circle. Most significant mean motion resonances
between the two innermost planets are labeled.}
\label{fig:stable}
\end{figure}

The dynamical analysis reveals two narrow strips of dominant solutions
around $a_d = 0.89$ AU (``Fit 1'', $\Chi = 1.30$ with rms 9.55
ms$^{-1}$) and 1.02~AU (``Fit 2'', $\Chi = 1.32$ with rms 9.58
ms$^{-1}$) (see Fig.\ \ref{fig:fit}), with some hint of a solution
around 2~AU. These are the Newtonian fits to the data. Table 1
presents these two fits, where $m$ is mass, $M$ is mean anomaly and
$T_0$ is the epoch of the first observation, JD~2,445,823.5570. To be
certain that the MEGNO signature is well time-calibrated (in the sense
that it will identify instabilities generated by strong mean motion
resonances), we integrated the best 16 solutions with the
Bulirsh-Stoer integrator from the MERCURY package (Chambers 1999) over
100~Myr each. All survived without any noticeable change of the
regular character. Fig.\
\ref{fig:stable} shows stable and chaotic regions near the two best
stable fits.

\begin{center}Table 1 - Best-fits Orbits and Masses for the Planets of HD 74156
\end{center}
\begin{center}
\begin{tabular}{ccccccc}
\hline\hline
Fit & Planet & $m$ (M$_{Jup}$) & $a$ (AU) & $e$ & $\omega$ ($^\circ$) & $M(T_0)$ ($^\circ$)\\
\hline
1 &  b  & 1.847 & 0.292 & 0.635 & 175.32 & 210.90 \\
 &  d  & 0.396 & 0.892 & 0.240 & 226.50 & 334.00 \\
 &  c  & 7.774 & 3.822 & 0.361 & 272.66 & 328.06 \\
\hline
2 & b & 1.847 & 0.292 & 0.629 & 176.45 & 211.64\\
 & d & 0.412 & 1.023 & 0.227 & 191.81 & 67.51\\
 & c & 7.995 & 3.848 & 0.426 & 262.17 & 340.20\\
\hline
3$^a$ & b & 1.882 & 0.292 & 0.640 & 175.37 & 210.88\\
 & d & 0.407 & 0.893 & 0.281 & 228.76 & 334.81\\
 & c & 7.830 & 3.818 & 0.363 & 272.364 & 327.39\\
\hline
4$^a$ & b & 1.882 & 0.292 & 0.635 & 176.20 & 211.55\\
 & d & 0.409 & 1.022 & 0.262 & 177.66 & 78.73\\
 & c & 8.176 & 3.853 & 0.427 & 261.25 & 341.43\\
\end{tabular}
\end{center}
$^a$ All planets in this fit have an inclination of $80^\circ$.\\

Finally, we tried to estimate a limit for the inclination of the
three-planet system. We selected a sample of 200 Newtonian solutions
from the previous search.  For each fit we increased the system's
inclination in steps of $1^{\circ}$ and then refined this fit with the
Levenberg-Marquart scheme.  Simultaneously, we also tested stability
of these fits with MEGNO.  Obviously, the inclination of the system is
unconstrained. Overall, the $N$-body model does not improve the formal
$\Chi$ of Keplerian -- but it is important for the dynamical analysis
and for finding stable configurations. Moreover, the stable fits can
be found down to $i \sim 40^{\circ}$ with simultaneous rough scaling
with the masses by the Doppler factor $f = 1/\sin i$. We also note
that the border of stable motions for $e_{\idm{d}}$ is at $\sim
0.3$. Still, fits with $a_{\idm{d}}\sim 0.89$~AU (``Fit 3'', $\Chi =
1.30$ with rms 9.6 ms$^{-1}$) yield a slightly smaller $\Chi$ than
solutions with $a_{\idm{d}}\sim 1.02$~AU (``Fit 4'', $\chi^2 = 1.31$
with rms 9.66 ms$^{-1}$). 

RB05 predicted a Saturn-mass planet was most likely to be discovered
in the range $0.9 \lsim a \lsim 1.4$ AU, and $0 \le e \lsim 0.2$. The
four most likely fits identified by this analysis show that the
prediction of the PPS model has been borne out. HD 74156 d is the
first extrasolar planet to have its mass and orbit predicted. Fits 1
and 3 have the same $\Chi$ value, but Fit 1 has a slightly smaller
rms, so it should be considered the best overall fit to the
observations.

\section{Discussion}

In addition to HD 74156, similar predictions of new planets had been
made for the systems 55 Cnc, HD 37124 and HD 38529 by BR04 and
RB05. Subsequently, the orbital architectures of three of these four systems have been updated in ways that are consistent with the PPS picture.

A fifth planet was recently discovered in orbit about 55 Cnc in the
stable zone identified in BR04 and RB05 (Fischer \etal 2008). This new
planet, 55 Cnc f, is at the inner edge of the stable zone, and is
consequently packed in with the inner three. 55 Cnc f is therefore
also consistent with the PPS model. However, there is still room for
additional planets in this system (Raymond \& Barnes 2008). The HD
37124 planetary system was revised from two to three planets, on
orbits markedly different from the initially-announced configuration
(Vogt \etal 2005; see also Go\'zdziewski \etal 2008). The current
best-fit orbits of the planets in HD 37124 (Go\'zdziewski \etal 2008)
suggest the system is packed and is therefore also consistent with the
PPS model. The final system considered in BR04 and RB05, HD 38529, has
yet to produce another planet, but the observations presented in
Moro-Mart\'in \etal (2007) suggest there is no observational evidence
for or against an additional planet in HD 38529.

The nature of the single planet systems appears at odds with the PPS
model.  About 85\% of known exoplanets are currently observed to be
the sole companion of their host
star\footnote{http://exoplanets.org}. If the PPS hypothesis is to be
believed, then we would naturally expect planetary systems should all
be multiple. There are three possible explanations for the current
observations of single planet systems in the context of the PPS
hypothesis: 1) The orbits of additional companions have not been
robustly detected yet, 2) the additional planets are too small to be
detected, and/or 3) the additional planets have significant
inclinations. Wright \etal (2007) find that at least 30\% of single
planet systems show evidence of additional companions. Therefore it
may be that, as more data are obtained, the fraction of single planet
systems will drop precipitously. Another possibility stems from the
observed distribution of planet masses, which rises steeply at low
mass (Marcy \etal 2005). This result implies many planets may lie in
these gaps, but surveys lack the precision to detect them. A third
possibility is that mutual inclinations could be significant,
resulting in only one detectable planet. For example, Marzari \&
Weidenschilling (2002) showed that scattering can pump up inclinations
to large values while simultaneously increasing
eccentricities. Therefore it may be that some systems appear to have
only one planet because the other planets are too inclined to the line
of sight to be detected by radial velocity surveys.

In addition to packed exoplanet systems, numerical experiments testing
the stability of the giant planets in our Solar System suggest that
they, too, are packed (\eg Varadi \etal 1999, Michtchenko \& Ferraz-Mello 2001; Barnes \& Quinn
2004). The apparently high frequency of packed giant planet systems
naturally motivates research to identify more planets in stable
locations. Stability analyses show that several systems stand
out as non-packed. A simple method to determine the proximity to
dynamical instability, $\beta$, is described in Barnes \& Greenberg
(2007) based on the concept of ``Hill stability'' (Marchal \& Bozis
1982; Gladman 1993). In Barnes \& Greenberg's formulation, the
stability boundary lies at $\beta = 1$, and stable systems require
$\beta > 1$, except in the case of mean motion resonances (note that
numerical analyses of these systems do suggest that they, too, are packed
[Go\'zdziewski \& Maciejewski 2001; Barnes \& Quinn 2004; Barnes \&
Greenberg 2007]). Barnes \& Greenberg 2007 estimated that when $\beta
> 2$, then the system may contain an additional planet, but more work
is needed to verify this possibility.

Several two-planet systems are known with $\beta$ values greater than
2 (note that HD 38529 has a $\beta$ value of
2.07). HD 217107 (Vogt \etal 2005) has a $\beta$ value of 7.1, HD 68988
(Wright \etal 2007) has $\beta = 12.5$ and HD 187123 (Wright \etal
2007) has $\beta = 15.3$\footnote{see http://www.lpl.arizona.edu/$\sim$rory/research/xsp/dynamics/ for an up-to-date table of $\beta$ values.}. These systems have gaps that are probably large enough to
contain multiple additional companions. We also note, however, that
the outer companions of the latter two systems have orbits which are
poorly constrained. Additionally, the inner planets have been tidally
circularized (Rasio \etal 1996) and consequently may have formed with
significantly larger values of $e$ and $a$ (Jackson\etal 2008). It may
be many years before the known planets' orbits are measured with
significant precision, let alone unknown companions are detected.

\section{Conclusions}
The correct prediction of the mass and orbit of HD 74156 d is a major
step forward in our understanding of the nature of exoplanets. It also
represents the first successful prediction of the mass and orbit of a
planet since Neptune was predicted independently by LeVerrier and
Adams in the 1840s. Although their approach was markedly different
than that of BR04 and RB05, both predictions relied on the dynamical
properties of other planets in the system.

As mentioned in $\S$ 1, of the first 6 multiple extrasolar planet
systems detected, only HD 74156 appeared to not be dynamically
packed. The detection of planet d by Bean \etal (2008), and its
confirmation presented in $\S$ 2, now means that the first 6 multi-planet
systems to be detected were packed. Although there may be some bias
toward detecting packed systems because radial velocity surveys are
more likely to find planets close to their host star, HD 74156 d was a
difficult planet to detect because of its low mass. Therefore its
detection is a strong indicator that most, if not all, multiple planet
systems are dynamically packed. But, of course, future observations
will ultimately reject or confirm the PPS theory. We
encourage observers to focus on HD 38529, HD 217107, HD 68988 and HD
187123 in order to determine if additional companions lay between the
two that are known. Furthermore, additional companions need to be
detected in one-planet systems.

We also encourage future work to continue to explore mechanisms which
may lead to packed planetary systems. It could be that packing is a
natural consequence of planet formation (Laskar 2000; Scardigli
2007). One such model is the dynamical relaxation of planets in a
damped medium (Adams \& Laughlin 2003), but more work is needed to
verify this possibility.

The detection of HD 74156 d suggests that planet formation is an
efficient process and that planets may be common. We have suggested
several systems which are important litmus tests for the PPS
theory. We also encourage programs aimed at detecting astrometric
signals from planets that have been detected by radial velocity
surveys.

%\medskip
R.B. acknowledges support from NASA's PG\&G grant NNG05GH65G and TPFFS
grant 811073.02.07.01.15. K.G. is supported by the Polish Ministry of
Sciences and Education, Grant No. 1P03D-021-29. S.N.R. was supported
by an appointment to the NASA Postdoctoral Program at the University of
Colorado Astrobiology Center, administered by Oak Ridge Associated
Universities through a contract with NASA. We thank Barbara McArthur,
Jacob Bean, Fritz Benedict, Steven Soter, Thomas Quinn, Chris Laws and
Richard Greenberg for helpful discussions.

\references
Adams, F.C. \& Laughlin, G. 2003, Icarus, 163, 290\\
Barnes, R. \& Greenberg, R. 2007, ApJ, 665, L67\\
Barnes, R. \& Quinn, T.R. 2001, ApJ, 554, 884\\
Barnes, R. \& Quinn, T.R. 2004, ApJ, 611, 494\\
Barnes, R. \& Raymond, S.N. 2004 ApJ, 617, 569 (BR04)\\
Bean, J.L., McArthur, B.E., Benedict, G.F. \& Armstrong, A. 2008, ApJ, 672, 1202\\
Bois, E. Kiseleva-Eggleton, L., Rambaux, N. \& Pilat-Lohinger, E. 2003, ApJ, 598, 1312\\
Butler, R.P. \etal 1999, ApJ, 526, 916\\
Chambers, J., 1999, MNRAS, 304, 793\\
Charbonneau, P. 1995, ApJS, 101, 309\\
Cincotta, P.M., Giordano, C.M. \& Simo, C. 2003, Physica D Nonlinear Phenomenon, 182, 151\\
Cincotta, P.M. \& Sim\'o, C. 2000, A\&AS, 147, 205\\
Fischer, D. \etal 2002, ApJ, 564, 1028\\
Gladman, B. 1993, Icarus, 106, 247\\
Go\'zdziewski, K., Breiter, S. \& Borczyk, W. 2008, MNRAS, 383, 989\\
Go\'zdziewski, K., Maciejewski, A.~J.2002, ApJ, 563, L81.\\
Go\'zdziewski, K., Maciejewski, A.~J. \& Migaszewski, C. 2007, ApJ, 657, 546\\
Go\'zdziewski, K. \& Migaszewski, C. 2006, A\&A, 449, 1219\\
Grosser, M. 1962, \textit{The Discovery of Neptune}, Harvard UP\\
Jackson, B., Greenberg, R. \& Barnes, R. 2008, ApJ, 678, 1396\\
Laskar, J. 2000, PhRvL, 84, 3240\\
Laughlin, G. \& Chambers, J.E. 2001, ApJ, 551, L109\\
Marchal, C. \& Bozis, G. 1982, CeMDA, 26, 311\\
Marcy, G.W. \etal 2001a, ApJ, 555, 418\\
Marcy, G.W. \etal 2001b, ApJ, 556, 296\\
Marcy, G.W. \etal 2005, PThPS, 158, 24\\
Marzari, F. \& Weidenschilling, S. 2002, Icarus, 156, 570\\
Michtchenko, T.A. \& Ferraz-Mello, S. 2001, AJ, 122, 474\\
-----------. 2001, Icarus, 149, 357.\\
Naef, D. \etal 2004, A\&A, 414, 351\\
Rasio, F.A., Tout, C. A., Lubow, S. H., \& Livio, M. 1996, ApJ, 470, 1187\\
Raymond, S.N. \& Barnes, R. 2005, ApJ, 619, 549 (RB05)\\
Raymond, S.N. \& Barnes, R. 2008, ApJ, submitted\\
Raymond, S.N., Barnes, R. \& Kaib, N.A. 2006, ApJ, 644, 1223\\
Raymond, S.N., Quinn, T.R., \& Lunine, J.I. 2007, Astrobiology, 7, 66\\
Rivera, E.J. \& Lissauer, J.J. 2000, ApJ, 530, 454\\
Scardigli, F. 2007, FoPh, 37, 1278\\
Varadi, F. Ghil, M., \& Kaula, W. M. 1999, Icarus, 139, 286.\\
Vogt, S.S. \etal 2005, ApJ, 632,638\\
Wright, J.T. \etal 2007, ApJ, 657, 533
\end{document}